\documentclass[preprint]{aastex}

\newcommand{\yr}{{\rm\,yr}}
\newcommand{\au}{{\rm\,AU}}

\begin{document}

\title{Diversity and Origin of 2:1 Orbital Resonances in
       Extrasolar Planetary Systems}
\author{Man Hoi Lee}
\affil{Department of Physics, University of California,
       Santa Barbara, CA 93106}
% \email{mhlee@europa.physics.ucsb.edu}

\begin{abstract}
A diversity of 2:1 resonance configurations can be expected in
extrasolar planetary systems, and their geometry can provide
information about the origin of the resonances.
Assembly during planet formation by the differential migration of
planets due to planet-disk interaction is one scenario for the origin
of mean-motion resonances in extrasolar planetary systems.
The stable 2:1 resonance configurations that can be reached by
differential migration of planets with constant masses and initially
coplanar and nearly circular orbits are
(1) anti-symmetric configurations with the mean-motion resonance
variables $\theta_1 = \lambda_1 - 2 \lambda_2 + \varpi_1$ and
$\theta_2 = \lambda_1 - 2 \lambda_2 + \varpi_2$ (where $\lambda_j$ and
$\varpi_j$ are the mean longitudes and the longitudes of periapse)
librating about $0^\circ$ and $180^\circ$, respectively (as in the
Io-Europa pair),
(2) symmetric configurations with both $\theta_1$ and $\theta_2$
librating about $0^\circ$ (as in the GJ 876 system), and
(3) asymmetric configurations with $\theta_1$ and $\theta_2$ librating
about angles far from either $0^\circ$ or $180^\circ$.
There are, however, stable 2:1 resonance configurations with symmetric
($\theta_1 \approx \theta_2 \approx 0^\circ$), asymmetric, and
anti-symmetric ($\theta_1 \approx 180^\circ$ and $\theta_2 \approx
0^\circ$) librations that cannot be reached by differential migration
of planets with constant masses and initially coplanar and nearly
circular orbits.
If real systems with these configurations are ever found, their origin
would require (1) a change in the planetary mass ratio $m_1/m_2$
during migration, (2) a migration scenario involving inclination
resonances, or (3) multiple-planet scattering in crowded planetary
systems.
We find that the asymmetric configurations with large $e_2$ and the
$\theta_1 \approx 180^\circ$ and $\theta_2 \approx 0^\circ$
configurations have intersecting orbits and that the $\theta_1 \approx
\theta_2 \approx 0^\circ$ configurations with $e_1 > 0.714$ have
prograde periapse precessions.
\end{abstract}
% \keywords{celestial mechanics --- planetary systems ---
%           planets and satellites: general}

\section{INTRODUCTION}

Mean-motion resonances may be ubiquitous in extrasolar planetary
systems.
There are as many as three resonant pairs of planets in the $12$
multiple planet systems discovered to date.
It is well established that the two planets about the star GJ 876
discovered by \citet{mar01} are in 2:1 orbital resonance, with both
lowest order, eccentricity-type mean-motion resonance variables,
\begin{equation}
\theta_1 = \lambda_1 - 2 \lambda_2 + \varpi_1
\end{equation}
and
\begin{equation}
\theta_2 = \lambda_1 - 2 \lambda_2 + \varpi_2
\end{equation}
(where $\lambda_{1,2}$ are the mean longitudes of the inner and outer
planet, respectively, and $\varpi_{1,2}$ are the longitudes of
periapse), librating about $0^\circ$ \citep{lau01,riv01,lee02}.
The simultaneous librations of $\theta_1$ and $\theta_2$,
and hence the secular apsidal resonance variable
\begin{equation}
\theta_3 = \varpi_1 - \varpi_2 = \theta_1 - \theta_2 ,
\end{equation}
about $0^\circ$ mean that the periapses are nearly aligned and that
conjunctions of the planets occur when both planets are near periapse.
The other two pairs of planets suspected to be in mean-motion
resonances are the pair about HD 82943 in 2:1 resonance \citep{may04}
and the inner two planets about 55 Cnc in 3:1 resonance \citep{mar02}.

The symmetric geometry of the 2:1 resonances in the GJ 876 system was
not expected, because the familiar 2:1 resonances between the Jovian
satellites Io and Europa are anti-symmetric, with $\theta_1$ librating
about $0^\circ$ but $\theta_2$ and $\theta_3$ librating about
$180^\circ$ (a configuration that would persist in the absence of
the Laplace resonance with Ganymede).
In the Io-Europa case, the periapses are nearly anti-aligned, and
conjunctions occur when Io is near periapse and Europa is near
apoapse.
\citet{lee02} have shown that the differences in the resonance
configurations are mainly due to the magnitudes of the orbital
eccentricities involved (see also \citealt{bm03}).
When the eccentricities are small, the Io-Europa configuration is
expected from the resonant perturbation theory to the lowest order in
the eccentricities.
However, the eccentricities of the GJ 876 system are sufficiently
large that there are large contributions from higher order terms, and
stable simultaneous librations of $\theta_1$ and $\theta_2$ require
$\theta_1 \approx \theta_2 \approx 0^\circ$ for a system with
eccentricities and masses like those in GJ 876.

A scenario for the origin of mean-motion resonances in extrasolar
planetary systems is that they were assembled during planet formation
by the differential migration of planets due to gravitational
interaction with the circumstellar disk from which the planets
formed.
A single giant planet (with a planet-to-star mass ratio $\ga$ a few
$\times 10^{-4}$) can open an annular gap in the circumstellar gas
disk about the planet's orbit.
If a disk forms two giant planets that are not separated too far (with
the ratio of the orbital semimajor axes $a_1/a_2 \sim 1/2$), the planets
can also clear the disk material between them rather quickly
\citep*{bry00,kle00,kle04}.
Disk material outside the outer planet exerts torques on the planet that
are not opposed by disk material on the inside, and the outer planet
migrates toward the star.
Any disk material left on the inside of the inner planet exerts
torques on the inner planet that push it away from the star.
The timescale on which the planets migrate is the disk viscous timescale,
whose inverse is \citep{war97}
\begin{eqnarray}
\left|{\dot a} \over a\right|
\approx {3 \nu \over 2 a^2}
&=& {3 \over 2} \alpha (H/a)^2 \Omega
\nonumber \\ & & \label{vistime}\\
&=& 9.4 \times 10^{-5} \left(\alpha \over 4\times 10^{-3}\right)
  \left(H/a \over 0.05\right)^2 P^{-1} ,
\nonumber
\end{eqnarray}
where ${\dot a} \equiv da/dt$, the kinematic viscosity $\nu$ is
expressed using the Shakura-Sunyaev $\alpha$ prescription ($\nu =
\alpha H^2 \Omega$), $H$ is the scale height of the disk, and $\Omega
= 2 \pi/P$ and $P \approx 2\pi a^{3/2}/(G m_0)^{1/2}$ are,
respectively, the mean motion and period of an orbit of semimajor axis
$a$.
Although the depletion of the inner disk means that the inner planet
may not move out very far, the condition of approaching orbits for
capture into mean-motion resonances is established.

\citet{lee02} have shown that the observed, symmetric, 2:1 resonance
configuration in the GJ 876 system can be easily established by the
differential migration due to planet-disk interaction (see also
\citealt*{sne01}).
They have also found that the observed eccentricities of the GJ 876
system require either significant eccentricity damping from
planet-disk interaction or resonance capture occurring just before
nebula dispersal, because continued migration of the planets while
locked in the 2:1 resonances can lead to rapid growth of the
eccentricities if there is no eccentricity damping.
As we shall see in \S 4, there are other types of stable 2:1
resonance configurations with large eccentricities for a system with
the inner-to-outer planet mass ratio, $m_1/m_2$, of the GJ 876 system,
but these configurations are either unstable for planets as massive as
those in GJ 876 or not reachable by differential migration of planets
with constant masses and coplanar orbits.

As \citet{lee02} pointed out, a practical way to investigate what
stable resonance configurations are possible from continued migration
of the planets after capture into 2:1 resonances is through numerical
migration calculations without eccentricity damping.
Either eccentricity damping or the termination of migration due to nebula
dispersal would lead to a system being left somewhere along a sequence.
They noted that asymmetric resonance configurations with $\theta_1$
and $\theta_2$ librating about angles other than $0^\circ$ and
$180^\circ$ are possible for systems with $m_1/m_2$ different from
those in GJ 876.
\citet{lee03a} have presented some results on the variety of stable
2:1 resonance configurations that can be reached by differential
migration (see also \citealt*{fbm03}).
Asymmetric librations were previously only known to exist for exterior
resonances in the restricted three-body problem (e.g.,
\citealt{mes58,bea94,mal99}).
Other recent works on 2:1 resonance configurations for two-planet
systems include those by \citet*{bfm03}, \citet{had03}, \citet{ji03},
and \citet{tho03}.
With the exception of \citet{tho03}, who examined the possibility of
inclination resonances for non-coplanar orbits, all of the works
mentioned in this paragraph have focused on systems with two planets
on coplanar orbits.

In this paper we present the results of a series of direct numerical
orbit integrations of planar two-planet systems designed to address
the following questions.
What are the possible planar 2:1 resonance configurations?
Can they all be obtained by differential migration of planets?
In \S 2 we describe the numerical methods.
In \S 3 we present the stable 2:1 resonance configurations that can be
reached by differential migration of planets with constant masses.
The analysis in \S 3 is an extension of those by \citet{lee03a} and
\citet{fbm03}.
We consider configurations with $0.1 \le m_1/m_2 \le 10$ and the
orbital eccentricity of the inner planet, $e_1$, up to $0.8$ and
provide a detailed analysis of properties such as the boundaries in
$m_1/m_2$ for different types of evolution, the region in parameter
space for configurations with intersecting orbits, and the region
where the retrograde periapse precessions expected for orbits in 2:1
resonances are reversed to prograde.
By using a combination of calculations in which $m_1/m_2$ is changed
slowly and differential migration calculations, we show in \S 4.1 that
there are also symmetric and asymmetric 2:1 resonance configurations
that cannot be reached by differential migration of planets with
constant masses.
In their studies of systems with masses similar to those in HD 82943,
\citet{ji03} and \citet{had03} have discovered anti-symmetric 2:1
resonance configurations with $\theta_1 \approx 180^\circ$ and
$\theta_2 \approx 0^\circ$.
We show in \S 4.2 that a sequence of this configuration with
anti-aligned, intersecting orbits exists for all $m_1/m_2$ examined.
In \S 5 we summarize our results and describe several mechanisms by
which the configurations found in \S 4 can be reached.

\section{NUMERICAL METHODS}

We consider a system consisting of a central star of mass $m_0$, an
inner planet of mass $m_1$, and an outer planet of mass $m_2$, with
the planets on coplanar orbits.
We refer to the planet with the smaller orbital semimajor axis as the
inner planet, but it should be noted that some of the configurations
studied below have intersecting orbits and the outer planet can be
closer to the star than the inner planet some of the time.
In addition to the mutual gravitational interactions of the star and
planets, the direct numerical orbit integrations presented below
include one of the following effects: forced orbital migration, a
change in $m_1/m_2$, or a change in $m_1 + m_2$.
We change $m_1/m_2$ according to $d\ln(m_1/m_2)/dt =$ constant and
keep $m_1 + m_2$ constant for the calculations with a change in
$m_1/m_2$, and vice versa for the calculations with a change in $m_1 +
m_2$.
For the calculations with forced orbital migration (and constant
masses), we force the outer planet to migrate inward with a migration
rate of the form ${\dot a}_2/a_2 \propto P_2^{-1}$.
The migration rate in equation (\ref{vistime}) is of this form if
$\alpha$ and $H/a$ are independent of $a$.
This form of the migration rate also has the convenient property that
the evolution of systems with different initial outer semimajor axis
$a_{2,0}$ but the same initial $a_1/a_2$ is independent of $a_{2,0}$
if we express the semimajor axes in units of $a_{2,0}$ and time in
units of the initial outer orbital period $P_{2,0}$.
Except for the migration calculations discussed in the last paragraph
of \S 3.2, there is no eccentricity damping.

The direct numerical orbit integrations were performed using modified
versions of the symplectic integrator SyMBA \citep*{dun98}.
SyMBA is based on a variant of the Wisdom-Holman (1991) method
and employs a multiple time step technique to handle close encounters.
The latter feature is essential for the integrations that involve
intersecting orbits and/or become unstable.
\citet{lee02} have described in detail how SyMBA can be modified to
include forced migration and eccentricity damping.
If there is no eccentricity damping, a single step of the modified
algorithm starts with changing $a_2$ according to the forced migration
term for half a time step, then evolving the system for a full time
step using the original SyMBA algorithm, and then changing $a_2$
according to the forced migration term for another half a time step.
For the calculations with a change in $m_1/m_2$ or $m_1 + m_2$, we
modify SyMBA to include the change in $m_1/m_2$ or $m_1 + m_2$ in the
same manner as the forced migration term.

Unlike the calculations in \citet{lee02}, which used astrocentric
orbital elements, we have input and output in Jacobi orbital elements,
and we apply the forced migration term to the Jacobi $a_2$ in the
differential migration calculations.
\citet{lee03b} have shown that Jacobi elements should be used in the
analysis of hierarchical planetary systems where $a_1/a_2$ is small
and that the use of astrocentric elements can introduce significant
high-frequency variations in orbital elements that should be nearly
constant on orbital timescales.
We have found through experiments that Jacobi elements are useful even
for 2:1 resonant systems with $a_1/a_2 \approx 2^{-2/3}$.
In their differential migration calculations for the GJ 876 system
with $(m_1 + m_2)/m_0 = 0.0095$, \citet{lee02} did not see the
libration of $\theta_2$ about $180^\circ$, which is expected for small
eccentricities.
However, when we repeated the calculations in Jacobi coordinates, we
were able to see the libration of $\theta_2$ about $180^\circ$ at
small eccentricities with the reduced fluctuations in the orbital
elements (see also \S 3.2 below).

\section{RESONANCE CONFIGURATIONS FROM DIFFERENTIAL MIGRATION OF
         PLANETS WITH CONSTANT MASSES}

\subsection{Results for $\mbox{\boldmath $(m_1 + m_2)/m_0 = 10^{-3}$}$
            and $\mbox{\boldmath ${\dot a}_2/a_2 = -10^{-6} /P_2$}$}

We begin with a series of differential migration calculations with
$m_1/m_2$ ranging from $0.1$ to $10$.
The total planetary mass $(m_1 + m_2)/m_0 = 10^{-3}$.
The planets are initially on circular orbits, with the ratio of the
semimajor axes $a_1/a_2 = 1/2$ (far from the 2:1 mean-motion
commensurability), and the outer planet is forced to migrate inward at
the rate ${\dot a}_2/a_2 = -10^{-6} /P_2$.
The small total planetary mass [compared to $(m_1 + m_2)/m_0 \approx
10^{-2}$ of the GJ 876 system] and slow migration rate (compared to
eq. [\ref{vistime}] with the nominal parameter values) are chosen to
reduce the libration amplitudes.
The slow migration rate is also chosen to reduce the offsets in the
centers of libration of the resonance variables due to the forced
migration.
The effects of a migration rate similar to that in equation
(\ref{vistime}) and a larger total planetary mass are discussed in
\S 3.2.

\begin{figure}[t]
\epsscale{0.42}
\plotone{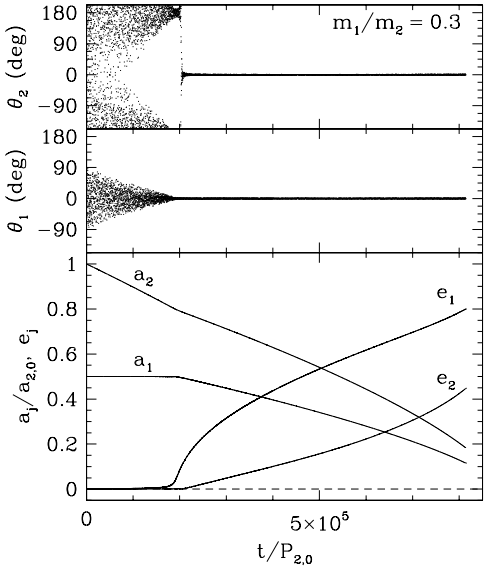}
\caption{
Evolution of the semimajor axes, $a_1$ and $a_2$, eccentricities,
$e_1$ and $e_2$, and 2:1 mean-motion resonance variables, $\theta_1 =
\lambda_1 - 2 \lambda_2 + \varpi_1$ and $\theta_2 = \lambda_1 - 2
\lambda_2 + \varpi_2$, for a differential migration calculation with
$m_1/m_2 = 0.3$ and $(m_1 + m_2)/m_0 = 10^{-3}$.
The outer planet is forced to migrate inward with ${\dot a}_2/a_2 =
-10^{-6} /P_2$.
The semimajor axes and time are in units of the initial orbital
semimajor axis, $a_{2,0}$, and period, $P_{2,0}$ of the outer planet,
respectively.
The sequence of resonance configurations after resonance capture is
$(\theta_1, \theta_2) \approx (0^\circ, 180^\circ) \rightarrow
(0^\circ, 0^\circ)$.
\label{figure1}
}
\end{figure}

\begin{figure}[t]
\epsscale{0.62}
\plotone{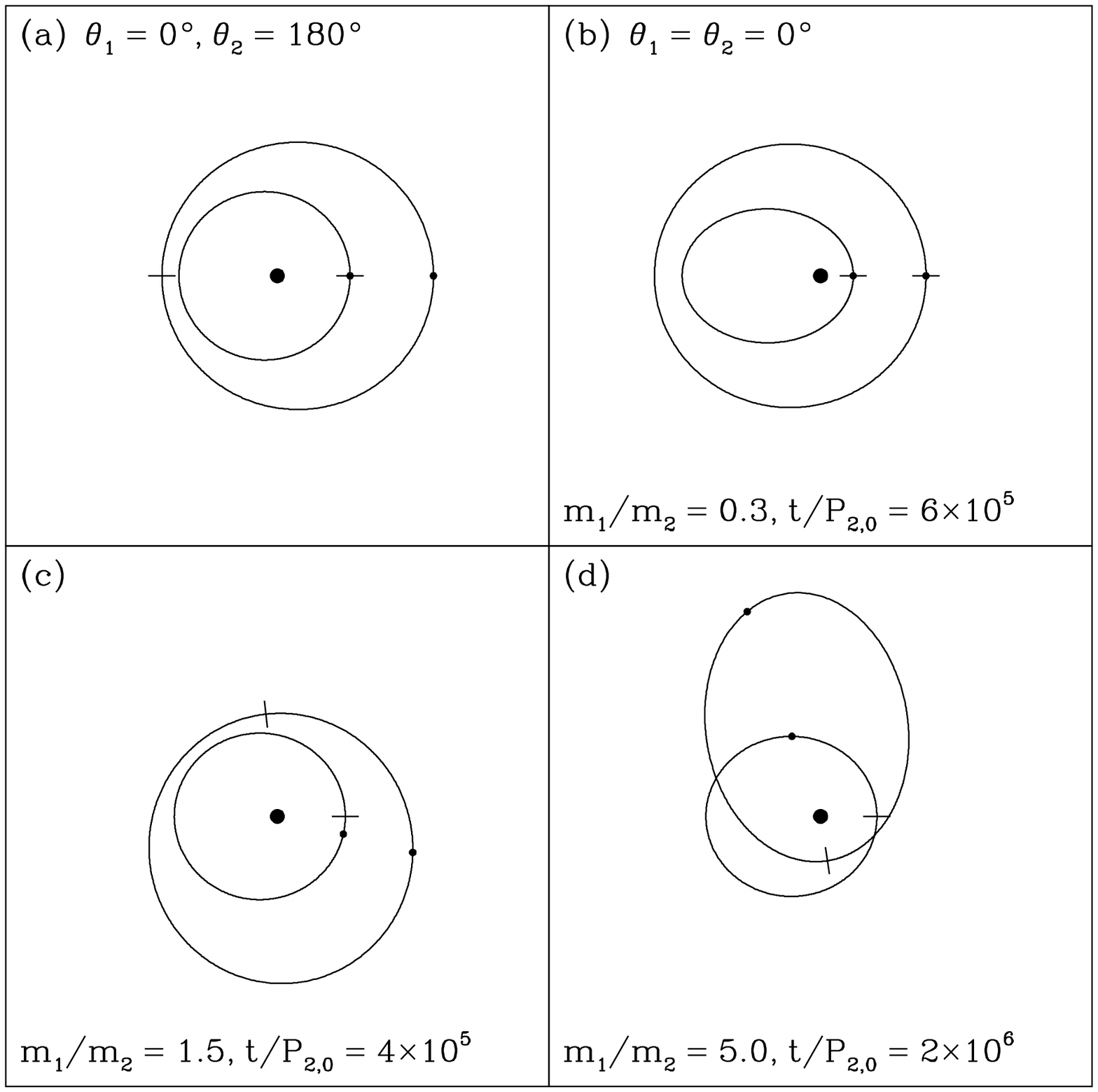}
\caption{
Examples of 2:1 resonance configurations that can be reached by
differential migration of planets with constant masses and initially
nearly circular orbits.
The dashes on the ellipses representing the orbits mark the positions
of the periapses, and the planets represented by the small dots are
shown at conjunction.
({\it a}) Anti-symmetric configuration with $\theta_1 \approx 0^\circ$
and $\theta_2 \approx 180^\circ$ at small eccentricities.
The eccentricities of the orbits are exaggerated so that the positions
of the periapses are more visible.
({\it b}) Symmetric configuration with $\theta_1 \approx \theta_2
\approx 0^\circ$ near $t/P_{2,0} = 6 \times 10^5$ in
Fig. \ref{figure1}.
({\it c}) Asymmetric configuration near $t/P_{2,0} = 4 \times 10^5$ in
Fig. \ref{figure3}.
({\it d}) Asymmetric configuration with intersecting orbits near
$t/P_{2,0} = 2 \times 10^6$ in Fig. \ref{figure4}.
\label{figure2}
}
\end{figure}

Figure \ref{figure1} shows the evolution of the semimajor axes $a_j$,
eccentricities $e_j$, and resonance variables $\theta_1$ and
$\theta_2$ for the calculation with $m_1/m_2 = 0.3$ (which is close to
$m_1/m_2 = 0.32$ of the GJ 876 system).
As the outer planet migrates inward, the 2:1 mean-motion
commensurability is encountered, and the system is first captured into
the 2:1 resonance configuration with $\theta_1$ librating about
$0^\circ$ and $\theta_2$ about $180^\circ$ (as in the Io-Europa pair).
Continued migration forces $e_1$ to larger values and $e_2$ from
increasing to decreasing until $e_2 \approx 0$ when $e_1 \approx 0.1$
(see Fig. \ref{figure5}{\it b} for a better view of this behavior).
Then the system passes smoothly over to the configuration with both
$\theta_1$ and $\theta_2$ librating about $0^\circ$ (as in the GJ 876
system), and both $e_1$ and $e_2$ increase smoothly.
The system remains stable for $e_1$ up to $0.8$, when the calculation
was terminated, although it may become unstable at larger $e_1$.
In Figure \ref{figure2}{\it a} we show the anti-symmetric resonance
configuration with $\theta_1 \approx 0^\circ$ and $\theta_2 \approx
180^\circ$.
The eccentricities of the ellipses representing the orbits are
exaggerated so that the positions of the periapses (marked by the
dashes) are more visible.
The planets (represented by the small dots) are shown at conjunction.
The periapses are nearly anti-aligned, and conjunctions occur when the
inner planet is near periapse and the outer planet is near apoapse.
In Figure \ref{figure2}{\it b} we show the symmetric resonance
configuration with $\theta_1 \approx \theta_2 \approx 0^\circ$.
The eccentricities are for the configuration near $t/P_{2,0} = 6
\times 10^5$ in Figure \ref{figure1}.
The periapses are nearly aligned, and conjunctions occur when both
planets are near periapse.
The evolution shown in Figure \ref{figure1} is characteristic for
$m_1/m_2 \la 0.95$.

\begin{figure}
\epsscale{0.42}
\plotone{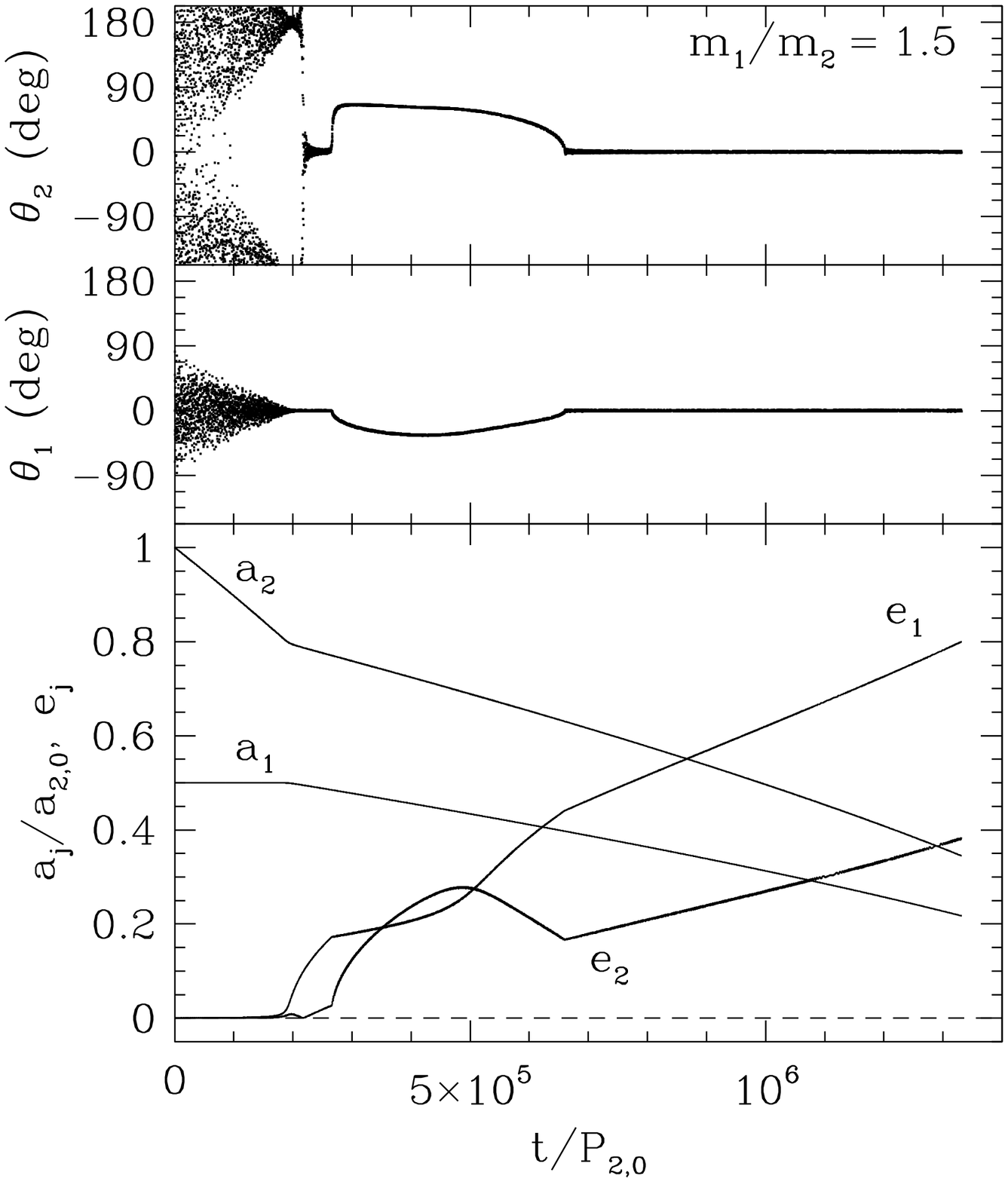}
\caption{
Same as Fig. \ref{figure1}, but for $m_1/m_2 = 1.5$.
The sequence of resonance configurations after resonance capture is
$(\theta_1, \theta_2) \approx (0^\circ, 180^\circ) \rightarrow
(0^\circ, 0^\circ) \rightarrow$ asymmetric librations $\rightarrow
(0^\circ, 0^\circ)$.
\label{figure3}
}
\end{figure}

\begin{figure}
\epsscale{0.42}
\plotone{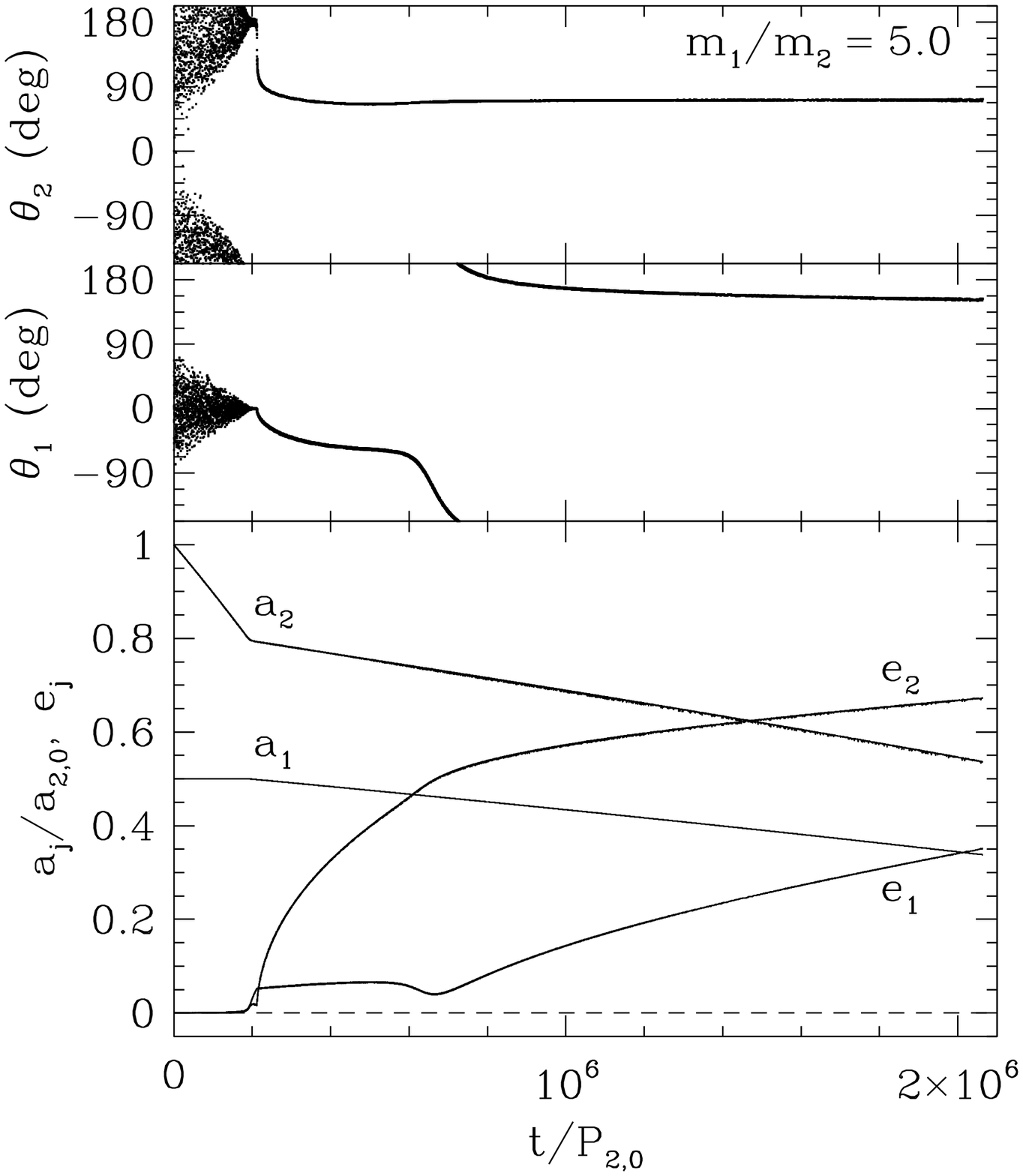}
\caption{
Same as Fig. \ref{figure1}, but for $m_1/m_2 = 5$.
The sequence of resonance configurations after resonance capture is
$(\theta_1, \theta_2) \approx (0^\circ, 180^\circ) \rightarrow$
asymmetric librations.
\label{figure4}
}
\end{figure}

Figure \ref{figure3} shows the evolution for the calculation with
$m_1/m_2 = 1.5$.
The capture into the $\theta_1 \approx 0^\circ$ and $\theta_2 \approx
180^\circ$ configuration and the passage to the $\theta_1 \approx
\theta_2 \approx 0^\circ$ configuration are as for $m_1/m_2 = 0.3$ in
Figure \ref{figure1}.
However, when the increasing $e_1$ reaches $\approx 0.17$, the
libration centers depart by tens of degrees from $0^\circ$, leading to
stable libration of the $\theta_j$ far from either $0^\circ$ or
$180^\circ$.
Asymmetric librations persist for $0.17 \la e_1 \la 0.44$, while $e_2$
changes from increasing to decreasing with increasing $e_1$.
The system returns to the configuration with both $\theta_1$ and
$\theta_2$ librating about $0^\circ$ and both $e_1$ and $e_2$
increasing smoothly for sufficiently large $e_1$($\ga 0.44$).
In Figure \ref{figure2}{\it c} we show the asymmetric resonance
configuration near $t/P_{2,0} = 4 \times 10^5$ in Figure
\ref{figure3}.
The angle between the lines of apsides is slightly more than
$90^\circ$, and conjunctions occur when neither planet is near
periapse or apoapse.
It should be noted that conjunction means equal {\it true} longitudes
of the planets, which is not the same as equal mean longitudes when
$\theta_1$ and $\theta_2$ are not $0^\circ$ or $180^\circ$.
The evolution shown in Figure \ref{figure3} is characteristic for
$0.95 \la m_1/m_2 \la 2.55$, with the minimum (maximum) $e_1$ for
asymmetric librations smaller (larger) for larger $m_1/m_2$ (see also
Fig. \ref{figure5}{\it a}).

There appears to be a narrow range in $m_1/m_2$ ($2.55 \la m_1/m_2 \la
2.75$) for which the system changes from $\theta_1 \approx 0^\circ$
and $\theta_2 \approx 180^\circ$ to asymmetric librations without
passing through the $\theta_1 \approx \theta_2 \approx 0^\circ$
configuration as $e_1$ increases, but the system does return to the
$\theta_1 \approx \theta_2 \approx 0^\circ$ configuration for
sufficiently large $e_1$.
Figure \ref{figure4} shows the evolution for the calculation with
$m_1/m_2 = 5$, which is characteristic for $m_1/m_2 \ga 2.75$.
The system changes from $\theta_1 \approx 0^\circ$ and $\theta_2
\approx 180^\circ$ to asymmetric librations and never reaches the
$\theta_1 \approx \theta_2 \approx 0^\circ$ configuration.
In the asymmetric libration branch, $e_2$ increases monotonically with
time but $e_1$ does not.
The system eventually becomes unstable at $t/P_{2,0} = 2.06 \times
10^6$.
The asymmetric libration configurations in Figure \ref{figure4} with
$e_2 \ga 0.35$ have the additional property that the orbits intersect.
An example --- the configuration near $t/P_{2,0} = 2 \times 10^6$ ---
is shown in Figure \ref{figure2}{\it d}.

It should be noted that the symmetries of the problem without forced
migration mean that every stable 2:1 resonance configuration with
centers of libration about $\theta_1$ and $\theta_2$ has a counterpart
with $-\theta_1$ and $-\theta_2$.
However, the forced migration causes small offsets in the libration
centers, and there could be a preference for evolution into one of the
asymmetric libration branches, depending on the migration rate and the
libration amplitudes.
As in the examples shown in Figures \ref{figure3} and \ref{figure4},
all of our calculations that yielded asymmetric configurations have
$\theta_1$ becoming negative when $\theta_1$ initially departs from
$0^\circ$.

\begin{figure}[t]
\epsscale{1.11}
\plottwo{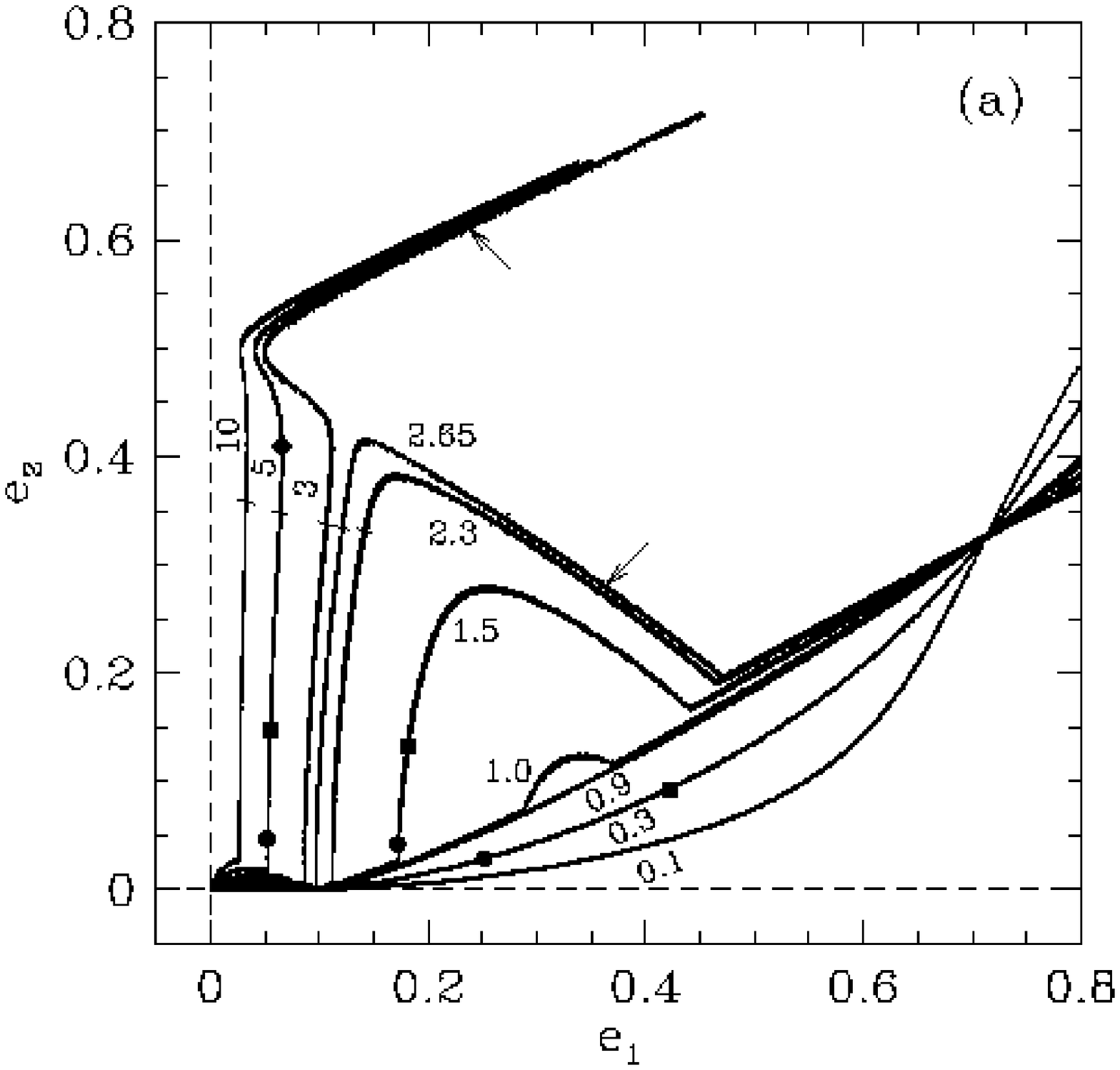}{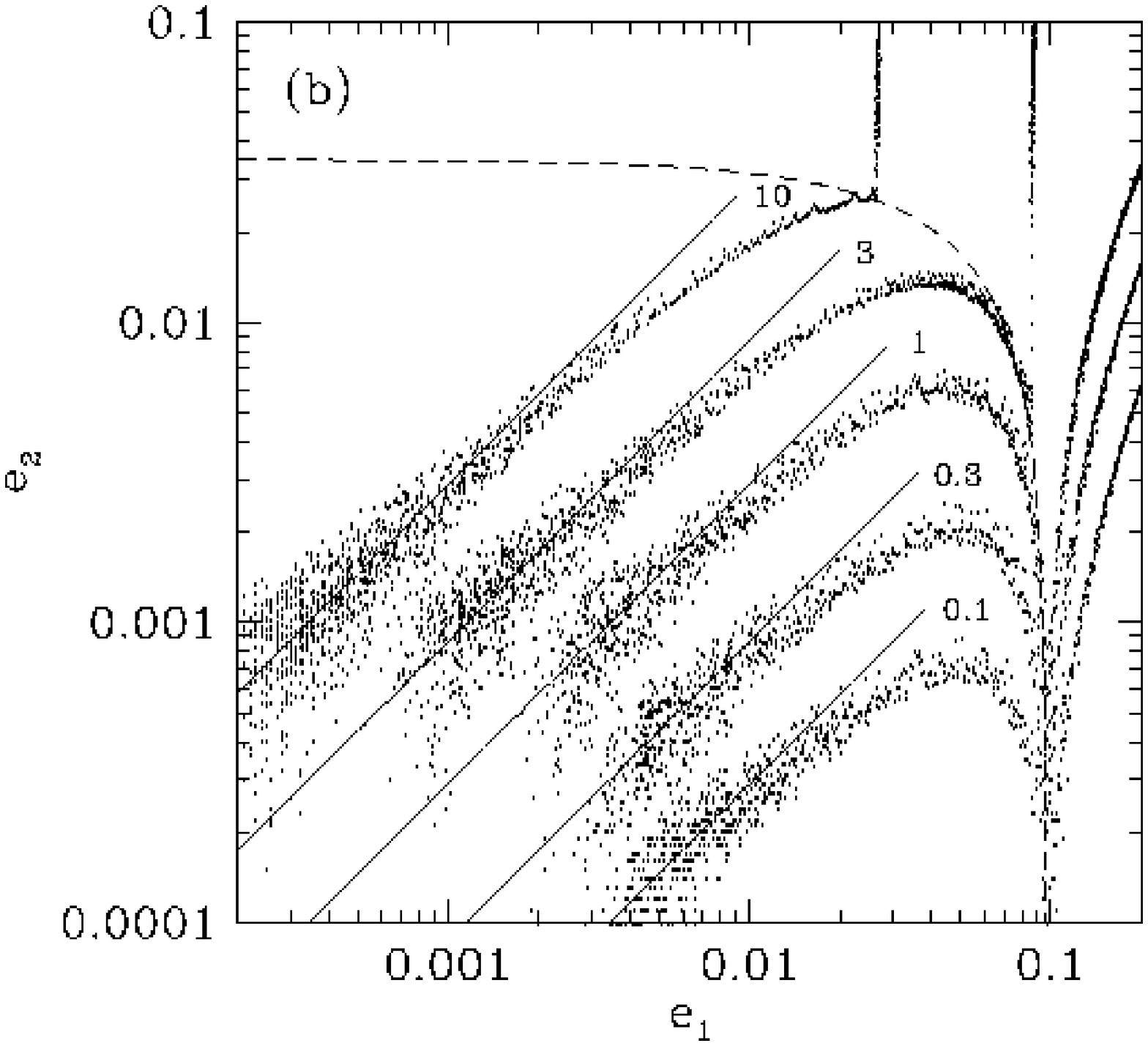}
\caption{
Trajectories of 2:1 resonance configurations from differential
migration of planets with constant masses and initially nearly
circular orbits.
The trajectories are labeled by $m_1/m_2$.
The total planetary mass $(m_1 + m_2)/m_0 = 10^{-3}$, and the outer
planet is forced to migrate inward with ${\dot a}_2/a_2 = -10^{-6}
/P_2$.
({\it a}) The $e_1$-$e_2$ plane in linear scales.
The configurations above the dash for $m_1/m_2 = 3$, $5$, and $10$ and
the configurations between the two dashes for $m_1/m_2 = 2.3$ and
$2.65$ have intersecting orbits.
The diamond, squares and circles on the curves with $m_1/m_2 = 0.3$,
$1.5$ and $5$ indicate the equilibrium eccentricities for eccentricity
damping of the outer planet with $K = |{\dot e}_2/e_2|/|{\dot
a}_2/a_2| = 1$, $10$ and $100$, respectively.
See text for $K = 1$ and $m_1/m_2 = 0.3$ and $1.5$.
The configurations indicated by the arrows on the curves with $m_1/m_2
= 2.65$ and $3$ are used as initial conditions for calculations in \S
3.2 and Figs. \ref{figure7} and \ref{figure8}.
({\it b}) The small eccentricity region of the $e_1$-$e_2$ plane in
logarithmic scales.
The solid lines show the relationship in eq. (\ref{e2e1}) from the
resonant perturbation theory to the lowest order in the
eccentricities.
The dashed line is the boundary below which the $\theta_1 = 0^\circ$
and $\theta_2 = 180^\circ$ configuration exists, as determined by
\citet{bfm03}.
\label{figure5}
}
\end{figure}

\begin{deluxetable}{ccl}
\tablecolumns{4}
\tablewidth{0pt}
\tablecaption{Sequences of 2:1 Resonance Configurations from
Differential Migration of Planets with Constant Masses
\label{table1}}
\tablehead{
\colhead{Planetary Mass Ratio} & \colhead{} & \colhead{Libration Centers $(\theta_1, \theta_2)$}
}
\startdata
$\phantom{0.95 \la}~m_1/m_2 \la 0.95$ & & $(0^\circ, 180^\circ) \rightarrow (0^\circ, 0^\circ)$ \\
$0.95 \la m_1/m_2 \la 2.55$ & & $(0^\circ, 180^\circ) \rightarrow (0^\circ, 0^\circ) \rightarrow$ asymmetric librations $\rightarrow (0^\circ, 0^\circ)$ \\
$2.55 \la m_1/m_2 \la 2.75$ & & $(0^\circ, 180^\circ) \rightarrow$ asymmetric librations $\rightarrow (0^\circ, 0^\circ)$ \\
$2.75 \la m_1/m_2~\phantom{\ga 2.75}$ & & $(0^\circ, 180^\circ) \rightarrow$ asymmetric librations \\
\enddata
\end{deluxetable}

A convenient and informative way to summarize the results for
different $m_1/m_2$ is to plot the trajectories in the $e_1$-$e_2$
plane (\citealt{bfm03}; \citealt{fbm03}).
Figure \ref{figure5} shows the trajectories for several values of
$m_1/m_2$ between $0.1$ and $10$, with Figure \ref{figure5}{\it a} in
linear scales and Figure \ref{figure5}{\it b} showing the small
eccentricity region in logarithmic scales.
The trajectories for $m_1/m_2 \la 0.95$, $0.95 \la m_1/m_2 \la 2.55$,
$2.55 \la m_1/m_2 \la 2.75$, and $m_1/m_2 \ga 2.75$ [with the
sequences of libration centers $(\theta_1, \theta_2)$ summarized in
Table \ref{table1}] are clearly distinguished in the $e_1$-$e_2$
plane.
The transition from $(\theta_1, \theta_2) \approx (0^\circ,
180^\circ)$ at small eccentricities to $(\theta_1, \theta_2) \approx
(0^\circ, 0^\circ)$ for $m_1/m_2 \la 2.55$ occurs when $e_1 \approx
0.1$ and $e_2 \approx 0$, and the transition to asymmetric librations
for $m_1/m_2 \ga 2.55$ occurs before $e_1$ reaches $0.1$, with the
critical value of $e_1$ smaller for larger $m_1/m_2$ (see
Fig. \ref{figure5}{\it b}).
\citet{bfm03} have determined the boundary in $e_1$-$e_2$ space below
which the $\theta_1 = 0^\circ$ and $\theta_2 = 180^\circ$
configuration exists, using an analytic model for the resonant
Hamiltonian (with an expansion in eccentricities).
Their result is well approximated by the relationship $e_2 = -0.36 e_1
+ 0.035$, which is shown as the dashed line in Figure
\ref{figure5}{\it b}.
It is in good agreement with the boundary in our numerical results.

For coplanar orbits, the equation for the variation of the periapse
longitude is
\begin{equation}
{d\varpi_j \over dt} = -{\sqrt{1 - e_j^2} \over m_j e_j \sqrt{G m_0 a_j}}\,
                        {\partial \Phi \over \partial e_j} ,
\label{dwdt}
\end{equation}
where the disturbing potential
\begin{equation}
\Phi = -{G m_1 m_2 \over a_2} \varphi (\beta, e_1, e_2, \theta_1, \theta_2) ,
\label{Phi}
\end{equation}
$\varphi$ is a function of the indicated variables, and $\beta =
a_1/a_2 \approx 2^{-2/3}$, if we consider only the resonant terms and
neglect terms of order $[(m_1 + m_2)/m_0]^2$ and higher (see, e.g.,
\citealt{lee02}; \citealt{bm03}).
To the lowest order in the eccentricities, $d\varpi_1/dt = \beta n_1
(m_2/m_0) C_1/e_1$ and $d\varpi_2/dt = -n_2 (m_1/m_0) C_2/e_2$ for
$\theta_1 = 0^\circ$ and $\theta_2 = 180^\circ$, where $n_j$ are the
mean motions and $C_1(\beta) = -1.19$ and $C_2(\beta) = +0.43$ are,
respectively, the coefficients of the $e_1 \cos\theta_1$ and $e_2
\cos\theta_2$ terms of the disturbing potential \citep{lee02}.
Thus, for 2:1 resonance configurations with sufficiently small
eccentricities, the precessions of the orbits are retrograde, and the
requirement that the orbits on average precess at the same rate
implies the following relationship between the forced eccentricities:
\begin{equation}
e_2/e_1 = -\beta^{1/2} (C_2/C_1) (m_1/m_2) = 0.29 (m_1/m_2) .
\label{e2e1}
\end{equation}
Our numerical results have finite widths in $e_1$ and $e_2$ due to
finite libration amplitudes, but they are on average in good agreement
with equation (\ref{e2e1}) (solid lines in Fig. \ref{figure5}{\it b})
for sufficiently small eccentricities.
We terminate the solid lines in Figure \ref{figure5}{\it b} at $e_1 =
0.12/(m_1/m_2 + 3)$, which is an estimate of where significant
deviation from equation (\ref{e2e1}) would occur, based on the
position of the local maximum of $e_2$ for $m_1/m_2 \ll 1$ and the
boundary in $e_1$-$e_2$ space for the $\theta_1 = 0^\circ$ and
$\theta_2 = 180^\circ$ configuration.

As we just mentioned, the orbital precessions are retrograde for the
resonance configurations with small eccentricities.
The precessions remain retrograde throughout the trajectories shown in
Figure \ref{figure5}{\it a} for $m_1/m_2 \ga 2.75$.
On the other hand, all of the trajectories for $m_1/m_2 \la 2.75$ pass
through a point with $e_1 = 0.714$ and $e_2 = 0.326$, where
$d\varpi_1/dt = d\varpi_2/dt = 0$ and the precessions of the orbits
change from retrograde to prograde.
(\citealt{lee02} have already noted the change from retrograde to
prograde precessions for systems with masses like those in GJ 876.)
According to equations (\ref{dwdt}) and (\ref{Phi}), the masses only
appear as overall factors in $d\varpi_j/dt$, with $d\varpi_1/dt
\propto m_2/\sqrt{m_0}$ and $d\varpi_2/dt \propto m_1/\sqrt{m_0}$.
Thus for small amplitude librations and nearly constant $e_j$ and
$\theta_j$, the combinations of $e_j$ and $\theta_j$ for which
$d\varpi_1/dt = d\varpi_2/dt$ is nonzero depend only on $m_1/m_2$,
and the combinations for which $d\varpi_1/dt = d\varpi_2/dt = 0$ are
independent of the planetary and stellar masses.
Since all of the cases in Figure \ref{figure5}{\it a} with $m_1/m_2
\la 2.75$ have $\theta_1$ and $\theta_2$ librating about $0^\circ$ at
large $e_1$, the only point on the $e_1$-$e_2$ plane where the
trajectories can intersect is a combination of $e_1$ and $e_2$ for
which $d\varpi_1/dt = d\varpi_2/dt = 0$.

In our discussion above of the case with $m_1/m_2 = 5$, we pointed out
that the asymmetric libration configurations with $e_2 \ga 0.35$ have
intersecting orbits (see, e.g., Fig. \ref{figure2}{\it d}).
\citet{fbm03} have stated that configurations with the apocentric
distance of the inner planet greater than the pericentric distance of
the outer planet [i.e., $a_1 (1 + e_1) > a_2 (1 - e_2)$ for $m_1, m_2
\ll m_0$] may have intersecting orbits, but this condition is
necessary but not sufficient (it is both necessary and sufficient only
for anti-aligned configurations such as those discussed in \S 4.2).
We have determined which of the configurations shown in Figure
\ref{figure5}{\it a} actually have intersecting orbits.
For $m_1/m_2 = 3$, $5$, and $10$, the configurations above the dash on
each trajectory have intersecting orbits.
However, configurations with intersecting orbits are not limited to
$m_1/m_2 \ga 2.75$.
For $m_1/m_2 = 2.3$ and $2.65$, the configurations between the two
dashes on each trajectory also have intersecting orbits.

\subsection{Effects of Migration Rate, Total Planetary Mass, and
            Eccentricity Damping}

\begin{figure}[t]
\epsscale{1.11}
\plottwo{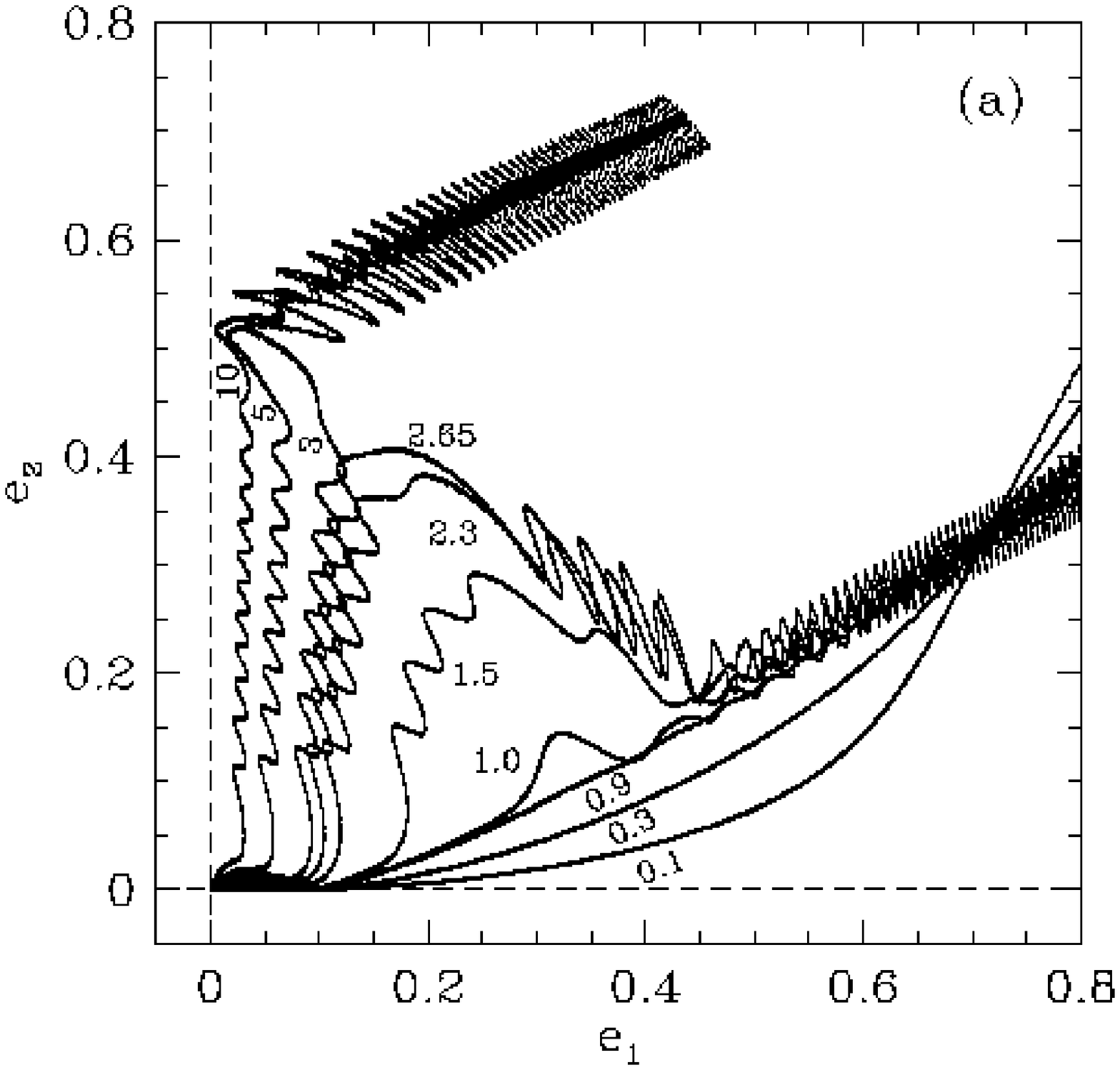}{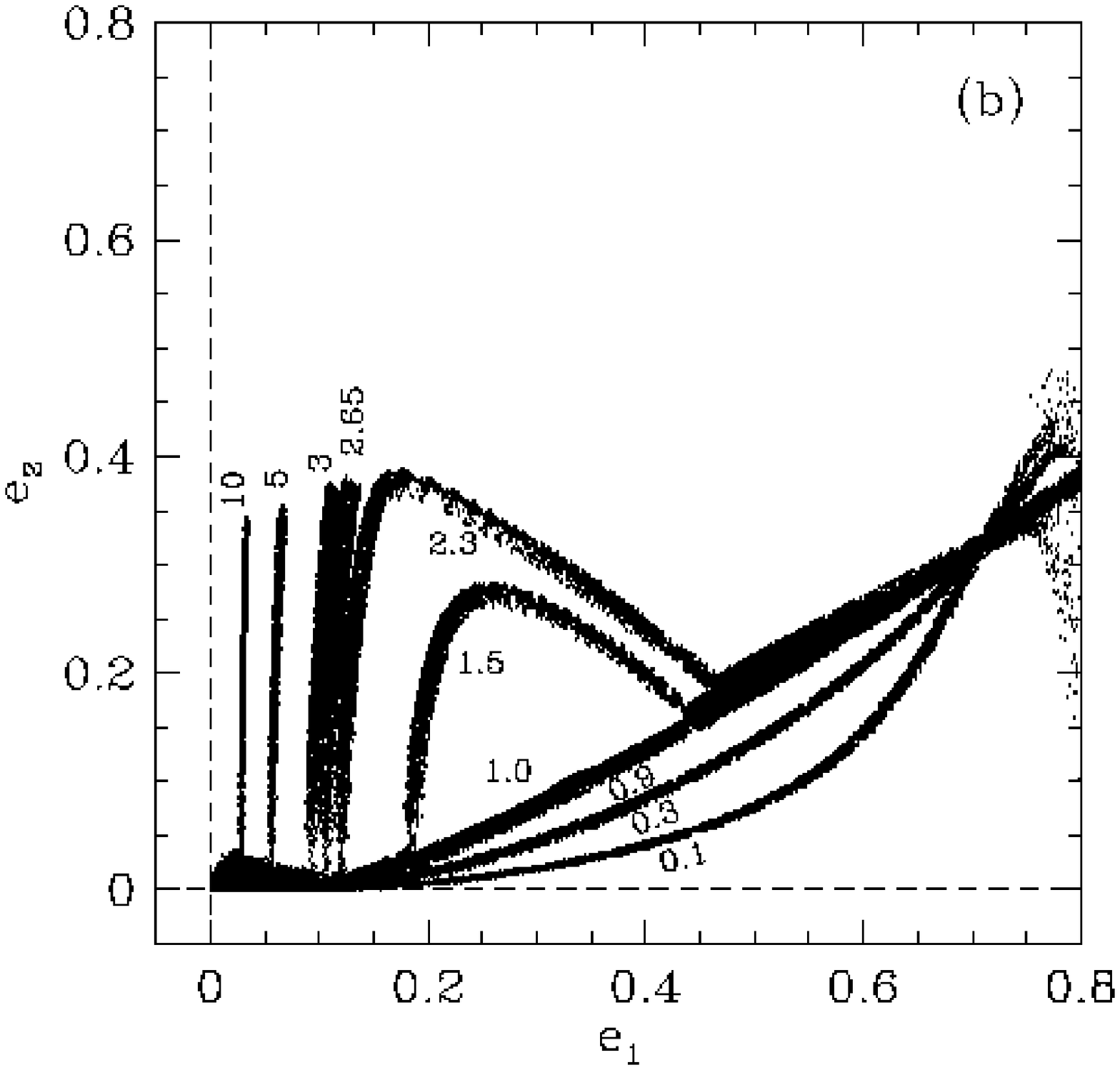}
\caption{
Same as Fig. \ref{figure5}{\it a}, but for ${\dot a}_2/a_2 =
-10^{-4}/P_2$ and ({\it a}) $(m_1 + m_2)/m_0 = 10^{-3}$ and ({\it b})
$(m_1 + m_2)/m_0 = 10^{-2}$.
The basic trajectories depend only on $m_1/m_2$, but the libration
amplitudes and the point at which a sequence becomes unstable can be
sensitive to the migration rate and the total planetary mass.
\label{figure6}
}
\end{figure}

We consider next the effects of a migration rate on the disk viscous
timescale similar to that in equation (\ref{vistime}) and a larger
total planetary mass.
We have performed differential migration calculations similar to those
in \S 3.1, but with ${\dot a}_2/a_2 = -10^{-4}/P_2$ and $(m_1 +
m_2)/m_0 = 10^{-3}$ and $10^{-2}$, and the trajectories of stable
resonance configurations in the $e_1$-$e_2$ plane are shown in Figures
\ref{figure6}{\it a} and \ref{figure6}{\it b}, respectively.
Figures \ref{figure5}{\it a} and \ref{figure6} show that the basic
trajectories depend only on $m_1/m_2$ and not on the total planetary
mass, in agreement with the analysis in \S 3.1 based on the
requirement of equal precession rates.
However, the libration amplitudes and the point at which a sequence
becomes unstable can be sensitive to the migration rate and the total
planetary mass.

For $(m_1 + m_2)/m_0 = 10^{-3}$, the faster migration rate
(Fig. \ref{figure6}{\it a}) leads to larger libration amplitudes
when a system with $m_1/m_2 \ga 0.95$ enters asymmetric libration, and
the point at which a system with $m_1/m_2 \ga 2.75$ becomes unstable
is slightly different from that shown in Figure \ref{figure5}{\it a}.
All of the trajectories in Figure \ref{figure6}{\it b} for $(m_1 +
m_2)/m_0 = 10^{-2}$ are wider than those in Figure
\ref{figure5}{\it a} because the larger eccentricity variations
generated by the larger total planetary mass before resonance capture
lead to resonance configurations with larger libration amplitudes.
On the other hand, although the trajectories in Figures
\ref{figure6}{\it a} and \ref{figure6}{\it b} are for the same
migration rate, those in Figure \ref{figure6}{\it b} with $m_1/m_2 \ga
0.95$ do not show any significant increase in the libration amplitudes
when a system enters asymmetric libration, because the larger total
planetary mass means that the resonant interaction between the planets
is stronger relative to the forced migration.
It is not clear in Figure \ref{figure6}{\it b}, but with the use of
Jacobi coordinates (see \S 2), we are able to see the capture into the
$\theta_1 \approx 0^\circ$ and $\theta_2 \approx 180^\circ$
configuration at small eccentricities when we examine time evolution
plots (like those in Figs. \ref{figure1}, \ref{figure3}, and
\ref{figure4}) for these calculations with $(m_1 + m_2)/m_0 =
10^{-2}$.

In Figure \ref{figure6}{\it b}, many of the trajectories for $m_1/m_2
\le 2.3$ show deviations from the trajectories in Figure
\ref{figure5}{\it a} at $e_1 \ga 0.75$ due to rapid increase in the
libration amplitudes before the system becomes unstable (see \S 4.1
for a more detailed discussion of a similar case).
The cases with $m_1/m_2 \ge 2.65$ become unstable near the points
where the orbits become intersecting (dashes on the trajectories in
Fig. \ref{figure5}{\it a}).
Note, however, that the case with $m_1/m_2 = 2.3$ passes through the
configurations with intersecting orbits without becoming unstable.
The fact that a sequence becomes unstable at a certain point does not
mean that all configurations beyond that point are necessarily
unstable, especially for smaller libration amplitudes.
We have used the asymmetric configuration, with non-intersecting
orbits, indicated by the arrow on the curve with $m_1/m_2 = 2.65$ in
Figure \ref{figure5}{\it a} as the starting point for a calculation
in which $m_1 + m_2$ is increased from $10^{-3} m_0$ at a rate of
$d\ln\left(m_1 + m_2\right)/dt = 10^{-6}/P_{2,0}$ while $m_1/m_2$ is
kept constant.
It remains stable to the end of the calculation when $(m_1 + m_2)/m_0
= 10^{-2}$.
When we used instead the asymmetric configuration, with intersecting
orbits, indicated by the arrow on the curve with $m_1/m_2 = 3$ in
Figure \ref{figure5}{\it a} as the starting point, the configuration
becomes unstable when $(m_1 + m_2)/m_0$ exceeds $0.0033$.

As we mentioned in \S 1, either eccentricity damping or the
termination of migration due to nebula dispersal would lead to a
system being left somewhere along a sequence in Figures \ref{figure5}
and \ref{figure6}.
There is significant uncertainty in both the sign and the magnitude of
the net effect of planet-disk interaction on the orbital eccentricity
of the planet because of sensitivity to the distribution of disk
material near the locations of the Lindblad and corotation resonances
(e.g., \citealt{gol03}).
Nevertheless, hydrodynamic simulations of two planets orbiting inside
an outer disk have shown eccentricity damping of the outer planet,
with $K = |{\dot e}_2/e_2|/|{\dot a}_2/a_2| \sim 1$ \citep{kle04},
while an explanation of the observed eccentricities of the GJ 876
system as equilibrium eccentricities requires $K \sim 100$
\citep{lee02}.
To study the effects of eccentricity damping, we have repeated the
$m_1/m_2 = 0.3$, $1.5$, and $5$ calculations shown in Figures
\ref{figure1}, \ref{figure3}, and \ref{figure4} with $K = 1$, $10$,
and $100$.
For $K = 10$ and $100$, the eccentricities reach equilibrium values in
all cases, and they are indicated by the squares and circles in
Figure \ref{figure5}{\it a}.
For $K = 1$, while the eccentricities in the $m_1/m_2 = 5$ calculation
reach equilibrium values indicated by the diamond in Figure
\ref{figure5}{\it a}, those in the $m_1/m_2 = 0.3$ and $1.5$
calculations are still increasing at the end of the calculations when
$e_1 \approx 0.68$ and $0.70$, respectively.

\section{OTHER RESONANCE CONFIGURATIONS}

\subsection{Symmetric and Asymmetric Configurations}

The bifurcation of the trajectories in Figure \ref{figure5}{\it a} at
$m_1/m_2 \approx 2.75$ leaves an empty region in the $e_1$-$e_2$
plane.
In this subsection we show that there are stable 2:1 resonance
configurations in the empty region of Figure \ref{figure5}{\it a} that
cannot reached by the scenario considered in \S 3 (i.e., differential
migration of planets with constant masses and initially nearly
circular orbits).

\begin{figure}
\epsscale{0.42}
\plotone{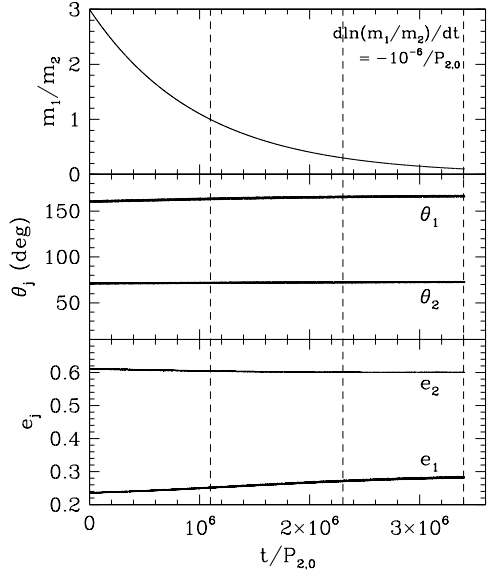}
\caption{
Evolution of the eccentricities, $e_1$ and $e_2$, mean-motion
resonance variables, $\theta_1$ and $\theta_2$, and mass ratio
$m_1/m_2$ for a calculation in which the configuration indicated by
the arrow on the curve with $m_1/m_2 = 3$ in Fig. \ref{figure5}{\it a}
is used as the starting point and $m_1/m_2$ is decreased at a rate of
$d \ln(m_1/m_2)/dt = -10^{-6}/P_{2,0}$.
The configurations with $m_1/m_2 = 1$, $0.3$, and $0.1$ indicated by
the dashed lines are used as initial conditions for calculations in
Figs. \ref{figure9} and \ref{figure11} and \S 4.3.
\label{figure7}
}
\end{figure}

\begin{figure}
\epsscale{0.42}
\plotone{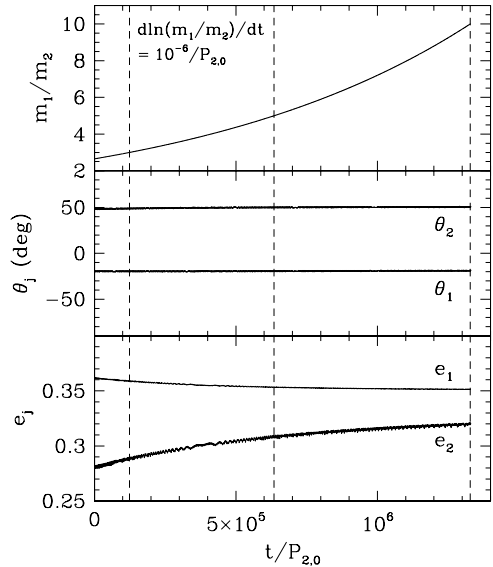}
\caption{
Same as Fig. \ref{figure7}, but for a calculation in which the
configuration indicated by the arrow on the curve with $m_1/m_2 =
2.65$ in Fig. \ref{figure5}{\it a} is used as the starting point and
$m_1/m_2$ is increased at a rate of $d \ln(m_1/m_2)/dt =
10^{-6}/P_{2,0}$.
The configurations with $m_1/m_2 = 3$, $5$, and $10$ indicated by
the dashed lines are used as initial conditions for calculations in
Figs. \ref{figure10} and \ref{figure11} and \S 4.3.
\label{figure8}
}
\end{figure}

In Figure \ref{figure7} we show the results of a calculation in which
the asymmetric configuration indicated by the arrow on the curve with
$m_1/m_2 = 3$ in Figure \ref{figure5}{\it a} is used as the starting
point and $m_1/m_2$ is decreased from 3 to 0.1 at a rate of $d
\ln(m_1/m_2)/dt = -10^{-6}/P_{2,0}$ while $(m_1 + m_2)/m_0 = 10^{-3}$
is kept constant.
As $m_1/m_2$ decreases, $\theta_1$ and $\theta_2$ remain near the
initial values, $e_2$ changes little, and $e_1$ increases (i.e., the
sequence extends to the right in the $e_1$-$e_2$ plane).
Similarly, we have used the asymmetric configuration indicated by the
arrow on the curve with $m_1/m_2 = 2.65$ in Figure \ref{figure5}{\it
a} as the starting point for a calculation in which $m_1/m_2$ is
increased from 2.65 to 10 at a rate of $d \ln(m_1/m_2)/dt =
10^{-6}/P_{2,0}$.
The results are shown in Figure \ref{figure8}.
As $m_1/m_2$ increases, $\theta_1$ and $\theta_2$ remain near the
initial values, $e_1$ changes little, and $e_2$ increases (i.e., the
sequence extends upward in the $e_1$-$e_2$ plane).
Thus we have found for each $m_1/m_2$ between $0.1$ and $10$ a
resonance configuration distinct from those shown in Figure
\ref{figure5}{\it a}.

\begin{figure}[t]
\epsscale{0.42}
\plotone{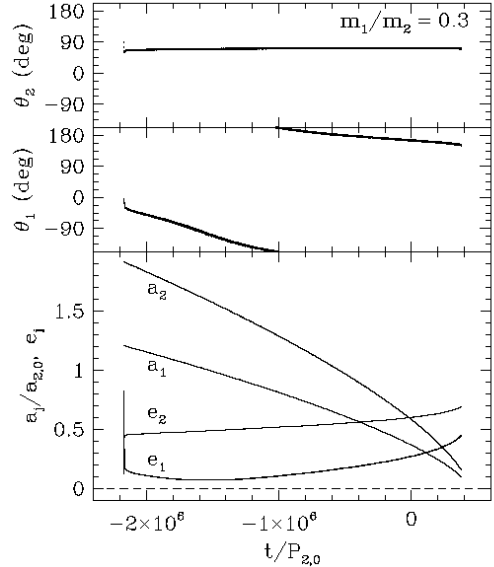}
\caption{
Resonance configurations from forward and backward migration
calculations using the $m_1/m_2 = 0.3$ configuration from
Fig. \ref{figure7} as initial conditions.
The results from the calculations with ${\dot a}_2/a_2 = -10^{-6}/P_2$
and $10^{-6}/P_2$ are plotted along the positive and negative
time axis, respectively.
These resonance configurations are different from those in
Fig. \ref{figure1} from differential migration of planets with
constant masses and initially nearly circular orbits.
\label{figure9}
}
\end{figure}

\begin{figure}[t]
\epsscale{0.42}
\plotone{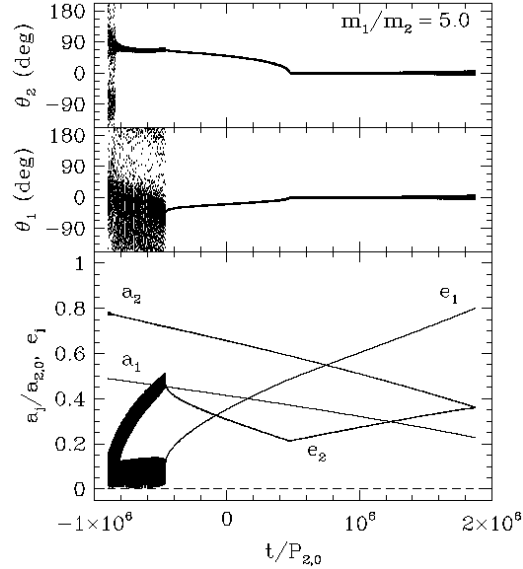}
\caption{
Same as Fig. \ref{figure9}, but for calculations using the $m_1/m_2 =
5$ configuration from Fig. \ref{figure8} as initial conditions.
The configurations at $t/P_{2,0} \ga -4.6 \times 10^5$ with both
$\theta_1$ and $\theta_2$ librating are different from those in
Fig. \ref{figure4}.
\label{figure10}
}
\end{figure}

\begin{figure}[t]
\epsscale{0.555}
\plotone{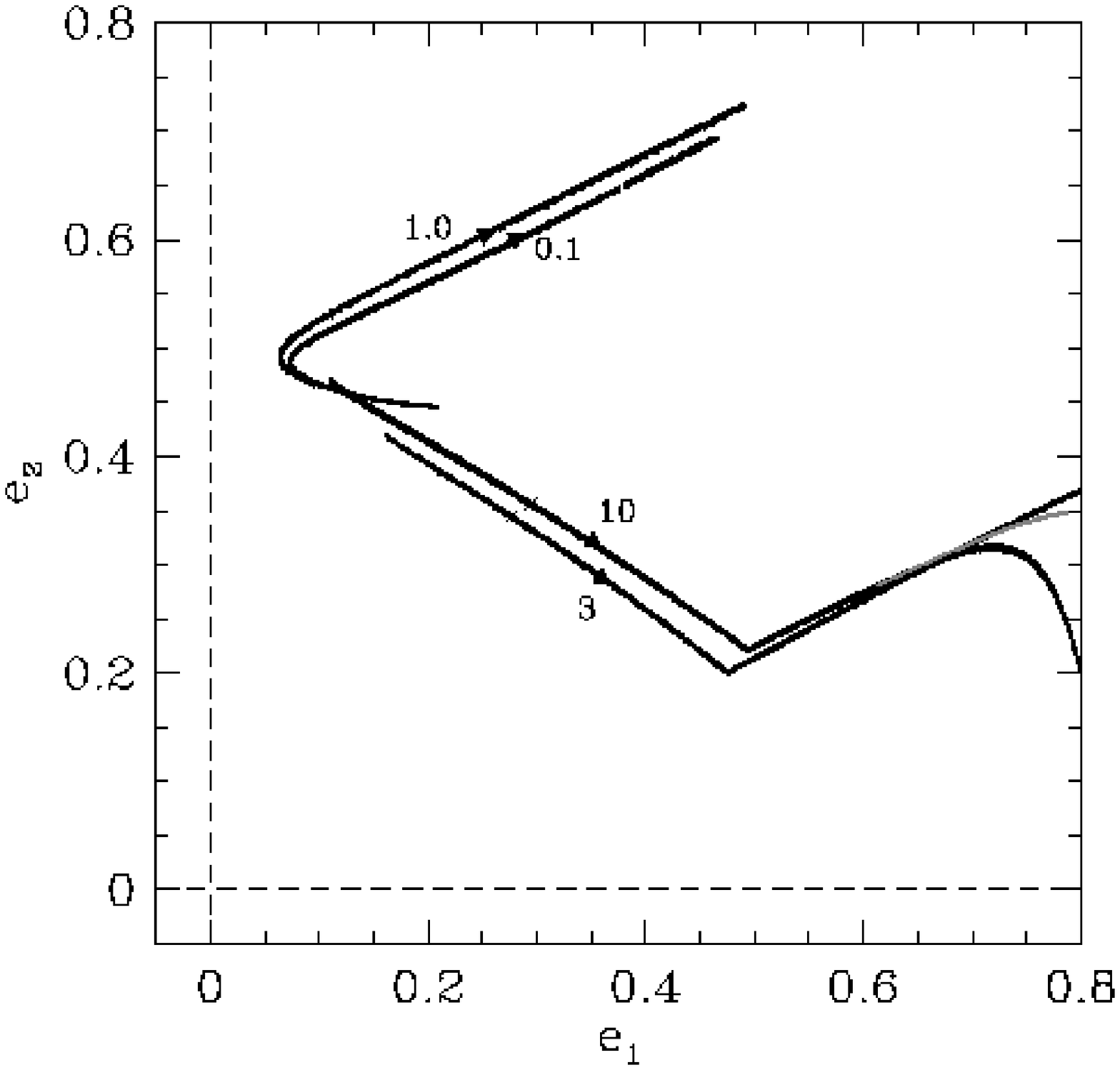}
\caption{
Trajectories in the $e_1$-$e_2$ plane of 2:1 resonance configurations
from forward and backward migration calculations with initial
conditions ({\it triangles}, oriented to indicate the direction for
forward migration) from Figs. \ref{figure7} and \ref{figure8} for
$m_1/m_2 = 0.1$, $1$, $3$, and $10$.
See text for the calculation that produces the gray curve for $m_1/m_2
= 10$.
The resonance configurations are distinct from those in
Fig. \ref{figure5}{\it a}.
All of the configurations for $m_1/m_2 = 0.1$ and $1$ and the
configurations to the left of the dashes on the trajectories for
$m_1/m_2 = 3$ and $10$ have intersecting orbits.
\label{figure11}
}
\end{figure}

For a given $m_1/m_2$, we can then use the resonance configuration
from either Figure \ref{figure7} (if $m_1/m_2 \la 2.75$) or Figure
\ref{figure8} (if $m_1/m_2 \ga 2.75$) as the starting point for
``forward'' (${\dot a}_2/a_2 < 0$) and ``backward'' (${\dot a}_2/a_2 >
0$) migration calculations to search for other resonance
configurations.
Figure \ref{figure9} shows the results from the $m_1/m_2 = 0.3$
calculations with ${\dot a}_2/a_2 = -10^{-6} /P_2$ and $10^{-6}/P_2$
along the positive and negative time axis, respectively.
In both calculations we have integrated the system until it becomes
unstable.
The asymmetric resonance configurations in Figure \ref{figure9} are
clearly different from the symmetric and anti-symmetric configurations
in Figure \ref{figure1}.
Figure \ref{figure10} shows the results from the $m_1/m_2 = 5$
calculations.
The calculation with ${\dot a}_2/a_2 = -10^{-6} /P_2$ was terminated
when $e_1$ reaches $0.8$, while the calculation with ${\dot a}_2/a_2 =
10^{-6} /P_2$ was terminated when both $\theta_1$ and $\theta_2$
circulate.
We shall not consider further configurations with only $\theta_2$
librating, but it is interesting that those in Figure \ref{figure10}
at $-8.5 \times 10^5 \la t/P_{2,0} \la -4.6 \times 10^5$ are similar
to those in Figure \ref{figure4} at $t/P_{2,0} \la 6 \times 10^5$ with
both $\theta_1$ and $\theta_2$ librating.
The configurations in Figure \ref{figure10} with both $\theta_1$ and
$\theta_2$ librating (at $t/P_{2,0} \ga -4.6 \times 10^5$) are either
symmetric or asymmetric, and they are clearly different from those in
Figure \ref{figure4}.

In Figure \ref{figure11} we show the trajectories in the $e_1$-$e_2$
plane of stable resonance configurations (with both $\theta_1$ and
$\theta_2$ librating) from the forward and backward
differential migration calculations with initial conditions (triangles
in Fig. \ref{figure11}) from Figures \ref{figure7} and \ref{figure8}
for $m_1/m_2 = 0.1$, $1$, $3$, and $10$.
The trajectories for $m_1/m_2 = 0.3$ and $5$, whose time evolutions
are shown in Figures \ref{figure9} and \ref{figure10}, respectively,
are simply between the appropriate pairs in Figure \ref{figure11} and
are not shown to avoid crowding.
Unlike the cases with $m_1/m_2 = 3$ and $5$ (see Fig. \ref{figure10}
for the latter), where $e_2$ increases monotonically with increasing
$e_1$ for the $\theta_1 \approx \theta_2 \approx 0^\circ$
configurations at large $e_1$, the trajectory for $m_1/m_2 = 10$ shows
a maximum in $e_2$ at $e_1 \approx 0.72$.
This turns out to be due to the libration amplitudes of
$\theta_1$ and $\theta_2$ becoming too large.
As the example in Figure \ref{figure10} shows, the libration amplitudes
of $\theta_1$ and $\theta_2$ increase as the system is driven deeper
into resonance.
The semi-amplitudes do not exceed $10^\circ$ in the cases with
$m_1/m_2 = 3$ and $5$, but for $m_1/m_2 = 10$, the configurations with
$e_1 \ga 0.68$ have semi-amplitudes of $\theta_1$ and $\theta_2 >
10^\circ$ (although $\theta_1$ and $\theta_2$ are nearly in phase so
that the libration amplitude of $\theta_3 = \theta_1 - \theta_2$
remains small and the trajectory in the $e_1$-$e_2$ plane remains
narrow).
We have repeated the $m_1/m_2 = 10$ calculation by starting at $e_1 =
0.61$ with initial conditions that correspond to smaller libration
amplitudes.
The resulting trajectory is shown as the gray curve in Figure
\ref{figure11}, and it does not show the decrease in $e_2$ at large
$e_1$.

Like the trajectories in Figure \ref{figure5}{\it a} for $m_1/m_2 \la
2.75$, the trajectories in Figure \ref{figure11} for $m_1/m_2 = 3$ and
$10$ (the gray curve for the latter) also pass through the point with
$e_1 = 0.714$ and $e_2 = 0.326$, where $d\varpi_1/dt = d\varpi_2/dt =
0$ and the precessions of the orbits change from retrograde at smaller
$e_1$ to prograde at larger $e_1$.
In Figure \ref{figure11}, all of the configurations for $m_1/m_2 =
0.1$ and $1$ and the configurations to the left of the dashes on the
trajectories for $m_1/m_2 = 3$ and $10$ have intersecting orbits.

The configurations shown in Figure \ref{figure11} clearly occupy a
different part of the $e_1$-$e_2$ plane compared to those in Figure
\ref{figure5}{\it a}.
They are found by a combination of calculations in which $m_1/m_2$ is
changed slowly and differential migration calculations, and cannot be
reached by differential migration of planets with constant masses and
initially nearly circular orbits.

\subsection{Anti-symmetric Configurations with
            $\mbox{\boldmath $\theta_1 \approx 180^\circ$}$ and
            $\mbox{\boldmath $\theta_2 \approx 0^\circ$}$}

\begin{figure}
\epsscale{0.31}
\plotone{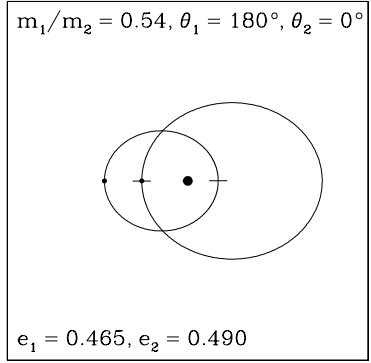}
\caption{
Anti-symmetric resonance configuration with $\theta_1 \approx
180^\circ$, $\theta_2 \approx 0^\circ$, and intersecting orbits used
as the starting point for calculations in Fig. \ref{figure13}.
\label{figure12}
}
\end{figure}

\begin{figure}
\epsscale{0.42}
\plotone{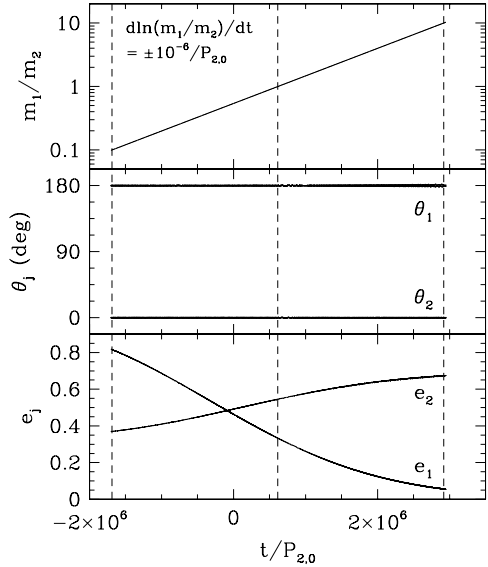}
\caption{
Evolution of the eccentricities, $e_1$ and $e_2$, mean-motion
resonance variables, $\theta_1$ and $\theta_2$, and mass ratio
$m_1/m_2$ for calculations in which the configuration shown in
Fig. \ref{figure12} is used as the starting point and $m_1/m_2$ is
increased to $10$ and decreased to $0.1$.
The results from the calculations with $d \ln(m_1/m_2)/dt =
10^{-6}/P_{2,0}$ and $-10^{-6}/P_{2,0}$ are plotted along the positive
and negative time axis, respectively.
The configurations with $m_1/m_2 = 0.1$, $1$, and $10$ indicated by
the dashed lines are used as initial conditions for calculations in
Fig. \ref{figure14}.
\label{figure13}
}
\end{figure}

As we mentioned in \S 1, \citet{ji03} and \citet{had03} have
discovered stable 2:1 resonance configurations with $\theta_1 \approx
180^\circ$ and $\theta_2 \approx 0^\circ$ for systems with masses
similar to those of the HD 82943 system announced initially by the
Geneva group [$m_1/m_2 = 0.54$, $(m_1 + m_2)/m_0 = 2.3 \times
10^{-3}$].
We have derived from the results of \citet{ji03} a $\theta_1 \approx
180^\circ$ and $\theta_2 \approx 0^\circ$ configuration with $m_1/m_2
= 0.54$, $(m_1 + m_2)/m_0 = 10^{-3}$, and small libration amplitudes,
and it is shown in Figure \ref{figure12}.
The $\theta_1 \approx 180^\circ$ and $\theta_2 \approx 0^\circ$
configuration has intersecting orbits, with the periapses nearly
anti-aligned and conjunctions occurring when the inner planet is near
apoapse and the outer planet is near periapse (and closer to the
star than the inner planet).

\begin{figure}[t]
\epsscale{0.555}
\plotone{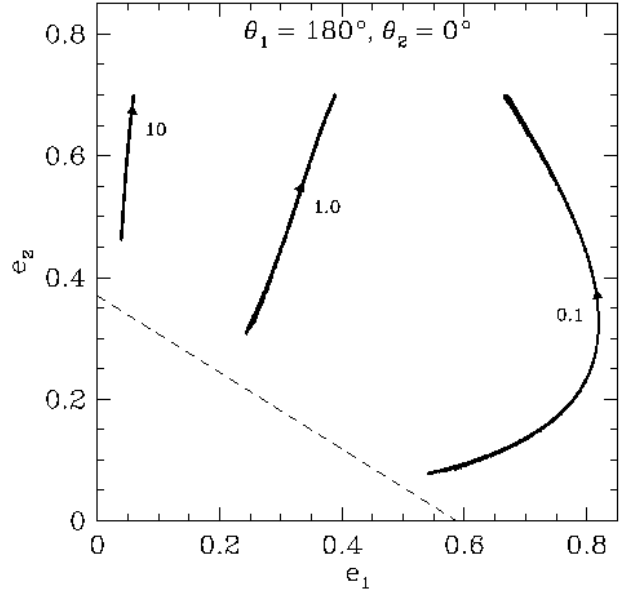}
\caption{
Trajectories in the $e_1$-$e_2$ plane of 2:1 resonance configurations
with $\theta_1 \approx 180^\circ$ and $\theta_2 \approx 0^\circ$ from
forward and backward migration calculations with initial conditions
({\it triangles}, oriented to indicate the direction for forward
migration) from Fig. \ref{figure13} for $m_1/m_2 = 0.1$, $1$, and
$10$.
These resonance configurations are distinct from those in
Figs. \ref{figure5}{\it a} and \ref{figure11}.
The trajectories terminate above the dashed line, where $a_1 (1+e_1) =
a_2 (1-e_2)$ and the planets would collide at conjunction.
\label{figure14}
}
\end{figure}

We have used the configuration in Figure \ref{figure12} as the
starting point for calculations in which $m_1/m_2$ is increased to
$10$ and decreased to $0.1$.
The results are shown in Figure \ref{figure13}, with the calculations
with $d \ln(m_1/m_2)/dt = 10^{-6}/P_{2,0}$ and $-10^{-6}/P_{2,0}$
along the positive and negative time axis, respectively.
The libration centers remain at $\theta_1 = 180^\circ$ and $\theta_2 =
0^\circ$, and $e_1$ decreases ($e_2$ increases) with increasing
$m_1/m_2$.
We have thus found for each $m_1/m_2$ in the range $0.1 \le m_1/m_2
\le 10$ a resonance configuration with $\theta_1 \approx 180^\circ$
and $\theta_2 \approx 0^\circ$.
As in \S 4.1, we can then use the resonance configuration for a given
$m_1/m_2$ from Figure \ref{figure13} as the starting point for forward
and backward migration calculations to search for other resonance
configurations with the same $m_1/m_2$.
Figure \ref{figure14} shows the trajectories in the $e_1$-$e_2$
plane of stable resonance configurations from the migration
calculations with $m_1/m_2 = 0.1$, $1$, and $10$.
The initial conditions (triangles in Fig. \ref{figure14}) are
indicated by the dashed lines in Figure \ref{figure13}.
The calculations with ${\dot a}_2/a_2 = -10^{-6} /P_2$ were terminated
when $e_2$ reaches $0.7$, while the calculations with ${\dot a}_2/a_2
= 10^{-6} /P_2$ were terminated when the system becomes unstable.
All of the configurations shown in Figure \ref{figure14} have
libration centers at $\theta_1 = 180^\circ$ and $\theta_2 = 0^\circ$
and retrograde orbital precessions.
Along the sequence for a given $m_1/m_2$, the fractional distance
between the planets at conjunction, $(a_1/a_2) (1+e_1) - (1-e_2)$,
decreases with decreasing $e_2$, and the system becomes unstable at a
point above the dashed line in Figure \ref{figure14}, where $a_1
(1+e_1) = a_2 (1-e_2)$ and the planets would collide at conjunction.
We can see from Figure \ref{figure14} that the configurations with
$\theta_1 \approx 180^\circ$ and $\theta_2 \approx 0^\circ$ tend to a
sequence with $e_1 = 0$ and $e_2$ above $0.37$ in the limit $m_1/m_2
\to \infty$.
These are the pericentric libration configurations found by
\citet{bea94} for the exterior 2:1 resonance in the planar, circular,
restricted three-body problem.

\subsection{Effects of Total Planetary Mass}

We consider in this subsection the effects of a larger total planetary
mass on the stability of the resonance configurations found in \S 4.1
and \S 4.2 for $(m_1 + m_2)/m_0 = 10^{-3}$.
We have used the configurations indicated by the dashed lines in
Figure \ref{figure7} as initial conditions for calculations in which
$m_1 + m_2$ is increased at a rate of $d\ln\left(m_1 + m_2\right)/dt =
10^{-6}/P_{2,0}$ while $m_1/m_2$ is kept constant.
These configurations with $m_1/m_2 = 1$, $0.3$, and $0.1$ become
unstable when $(m_1 + m_2)/m_0$ exceeds $0.0037$, $0.0041$, and
$0.0043$, respectively.
Similar calculations with the configurations indicated by the dashed
lines in Figure \ref{figure8} with $m_1/m_2 = 3$, $5$, and $10$ as
initial conditions show that these configurations are stable for $(m_1
+ m_2)/m_0$ up to at least $10^{-2}$.

For the anti-symmetric configurations with $\theta_1 \approx
180^\circ$ and $\theta_2 \approx 0^\circ$, the minimum $e_2$ (and the
minimum fractional distance between the planets at conjunction) at
which the configurations become unstable increases with increasing
total planetary mass.
We have used the configurations at the top of the trajectories shown
in Figure \ref{figure14} with $e_2 \approx 0.7$ and $m_1/m_2 = 0.1$,
$1$, and $10$ as initial conditions for calculations in which $m_1 +
m_2$ is increased, and they remain stable to the end of the
calculations when $(m_1 + m_2)/m_0 = 10^{-2}$.
When we used instead the configurations indicated by the triangles in
Figure \ref{figure14} as initial conditions, the cases with $m_1/m_2 =
0.1$ and $1$ are stable for $(m_1 + m_2)/m_0$ up to at least
$10^{-2}$, but the case with $m_1/m_2 = 10$ becomes unstable when
$(m_1 + m_2)/m_0$ exceeds $0.0096$.

It is likely that many (if not all) of the asymmetric configurations
shown in Figure \ref{figure9} for $m_1/m_2 = 0.3$ (which is close to
$m_1/m_2 = 0.32$ of the GJ 876 planets) are unstable for the total
mass of the GJ 876 planets [$(m_1 + m_2)/m_0 = 0.0095$], because the
configuration at $t = 0$ is the $m_1/m_2 = 0.3$ configuration from
Figure \ref{figure7} examined above and it is unstable for $(m_1 +
m_2)/m_0 > 0.0041$.
Thus for a system with masses like those in GJ 876, there are stable
2:1 resonance configurations with $(\theta_1, \theta_2) \approx
(180^\circ, 0^\circ)$, in addition to those with $(\theta_1, \theta_2)
\approx (0^\circ, 180^\circ)$ and $(0^\circ, 0^\circ)$ that can be
obtained from differential migration of planets with constant masses.
However, the planets are sufficiently massive that asymmetric
configurations like those in Figure \ref{figure9} are probably
unstable.

\section{DISCUSSION AND CONCLUSIONS}

We have shown that there is a diversity of 2:1 resonance
configurations for planar two-planet systems.
We began with a series of differential migration calculations, with
planets having constant masses and initially circular orbits, and
found the following types of stable 2:1 resonance configurations:
(1) anti-symmetric configurations with the mean-motion resonance
variables $\theta_1 = \lambda_1 - 2 \lambda_2 + \varpi_1$ and
$\theta_2 = \lambda_1 - 2 \lambda_2 + \varpi_2$ librating about
$0^\circ$ and $180^\circ$, respectively (as in the Io-Europa pair);
(2) symmetric configurations with both $\theta_1$ and $\theta_2$
librating about $0^\circ$ (as in the GJ 876 system); and
(3) asymmetric configurations with $\theta_1$ and $\theta_2$ librating
about angles far from either $0^\circ$ or $180^\circ$.
Systems with $m_1/m_2 \la 0.95$, $0.95 \la m_1/m_2 \la 2.55$, $2.55
\la m_1/m_2 \la 2.75$, and $m_1/m_2 \ga 2.75$ show different types of
evolution (Table \ref{table1}), and their trajectories in the
$e_1$-$e_2$ plane are clearly distinguished (Fig. \ref{figure5}).
The basic trajectories depend only on $m_1/m_2$, but the libration
amplitudes and the point at which a sequence becomes unstable can be
sensitive to the migration rate and the total planetary mass.
Where a system is left along a sequence depends on the magnitude of
the eccentricity damping or the timing of the termination of migration
due to nebula dispersal.
There are also stable 2:1 resonance configurations with symmetric
($\theta_1 \approx \theta_2 \approx 0^\circ$), asymmetric, and
anti-symmetric ($\theta_1 \approx 180^\circ$ and $\theta_2 \approx
0^\circ$) librations that cannot be reached by differential migration
of planets with constant masses and initially nearly circular orbits.
We have analyzed the properties of the resonance configurations found
in our study.
The asymmetric configurations with large $e_2$ and the $\theta_1
\approx 180^\circ$ and $\theta_2 \approx 0^\circ$ configurations have
intersecting orbits, while the $\theta_1 \approx \theta_2 \approx
0^\circ$ configurations with $e_1 > 0.714$ have prograde orbital
precessions.
Figures \ref{figure5}, \ref{figure11}, and \ref{figure14} can be used
to determine whether an observed system could be near a 2:1 resonance
configuration with both $\theta_1$ and $\theta_2$ librating.

We now describe several mechanisms by which the resonance
configurations summarized in Figures \ref{figure11} and \ref{figure14}
can be reached.
Since we found the symmetric and asymmetric configurations in \S 4.1
(Fig. \ref{figure11}) by a combination of calculations in which
$m_1/m_2$ is changed and differential migration calculations, they can
be reached by differential migration if the planets continue to grow
at different rates and $m_1/m_2$ changes during migration.
In particular, if $m_1/m_2 \la 2.75$ when the system is first captured
into resonance and $m_1/m_2$ increases above $2.75$ when $e_1 \ga
0.17$, the system should cross over to the configurations with
$m_1/m_2 \ga 2.75$ in Figure \ref{figure11}.
Similarly, if $m_1/m_2 \ga 2.75$ when the system is first captured
into resonance and $m_1/m_2$ decreases below $2.75$ when $e_2 \ga
0.49$, the system should cross over to the configurations with
$m_1/m_2 \la 2.75$ in Figure \ref{figure11}.
After the planets open gaps about their orbits and clear the disk
material between them, they can continue to grow by accreting gas
flowing across their orbits from the inner and outer disks.
If the inner disk is depleted, the situation with the outer planet
growing faster and a decreasing $m_1/m_2$ may be more likely
\citep{kle00}.

A mechanism by which the $\theta_1 \approx 180^\circ$ and $\theta_2
\approx 0^\circ$ configuration in \S 4.2 (Fig. \ref{figure14}) can be
reached is a migration scenario involving inclination resonances.
\citet{tho03} have studied the capture into and evolution in 2:1
resonances due to differential migration for orbits with small initial
inclinations.
For the case with $m_1/m_2 = 1$, the system is initially captured into
eccentricity-type mean-motion resonances only (with both $\theta_1$
and $\theta_2$ librating), and the initial evolution after capture is
similar to that of the planar system discussed in \S 3.
But when $e_1$ grows to $\sim 0.6$, the system also enters into
inclination-type mean-motion resonances (with both $2 \lambda_1 - 4
\lambda_2 + 2 \Omega_1$ and $2 \lambda_1 - 4 \lambda_2 + 2 \Omega_2$
librating, where $\Omega_{1,2}$ are the longitudes of ascending node),
and the orbital inclinations begin to grow.
The system eventually evolves out of the inclination resonances, and
interestingly, the eccentricity resonances switch from the $\theta_1
\approx \theta_2 \approx 0^\circ$ configuration to the $\theta_1
\approx 180^\circ$ and $\theta_2 \approx 0^\circ$ configuration at the
same time.
\citet{tho03} have noted that the inclination resonances are not
encountered until $e_1$ grows to $\sim 0.6$, which requires $m_1/m_2$
to be less than about $2$ for planets with constant masses.
From the results in \S 3 (Fig. \ref{figure5}{\it a}), we find that the
boundary in $m_1/m_2$ for $e_1$ to reach $\sim 0.6$ is more precisely
$m_1/m_2 \la 2.75$.
However, if we allow $m_1/m_2$ to change during migration as described
in the previous paragraph, it would be possible for $e_1$ to grow to
$\sim 0.6$ even for systems with final $m_1/m_2 \ga 2.75$ (see
Fig. \ref{figure11}).
Thus it may be possible to reach the $\theta_1 \approx 180^\circ$ and
$\theta_2 \approx 0^\circ$ configuration by evolving through
inclination resonances for non-coplanar systems with a wide range of
final $m_1/m_2$.

Finally, there is a mechanism for establishing mean-motion resonances
in planetary systems that does not involve migration due to
planet-disk interaction.
\citet*{lev98} have found that a significant fraction of systems at
the end of a series of simulations starting with a large number of
planetary embryos has mean-motion resonances.
More recently, \citet{ada03} have performed a series of $N$-body
simulations of crowded planetary systems, with $10$ giant planets
initially in the radial range $5 \au \le a \le 30 \au$.
After $10^6 \yr$, there are often only two or three planets remaining
in a system, and a surprisingly large fraction ($\sim 10\%$) of the
systems have a pair of planets near resonance (roughly equally
distributed among the 2:1, 3:2, and 1:1 resonances), although not all
of the resonant pairs will survive on longer timescales.
If there is no additional damping mechanism, the resonance
configurations produced by multiple-planet scattering in crowded
planetary systems are likely to have moderate to large orbital
eccentricities and relatively large libration amplitudes.
A more detailed analysis of the resonances produced by this mechanism
is needed, but there is no obvious reason to suspect that it cannot
produce any of the 2:1 resonance configurations with moderate to large
orbital eccentricities presented in \S 3 and \S 4, provided that the
configuration is stable for relatively large libration amplitudes.

The number of multiple-planet systems is expected to increase
significantly in the next few years.
Our results show that a wide variety of 2:1 resonance configurations
can be expected among future discoveries and that their geometry can
provide information about the origin of the resonances.

\acknowledgments
It is a pleasure to thank Stan Peale for his contributions to the
early stages of this work and for many useful discussions.
I also thank G. Laughlin and E. W. Thommes for informative discussions.
This research was supported in part by NASA grants NAG5-11666 and
NAG5-13149.

\end{document}